\documentclass[a4paper,twocolumn,preprint,3p]{elsarticle}

\usepackage{lineno,hyperref}

\modulolinenumbers[5]

\journal{Physica E. }

\begin{document}

\begin{frontmatter}

\title{Shot noise fluctuations in disordered graphene nanoribbons near the Dirac point}

\author{V\'ictor A. Gopar}
\address{Departamento de F\'isica Te\'orica and BIFI, Universidad de Zaragoza, Pedro Cerbuna 12, E-50009, Zaragoza, Spain.}
\fntext[myfootnote]{Special issue ``\it{Frontiers in 
Quantum Electronic Transport-In Memory of Markus B\"uttiker}''}

\cortext[mycorrespondingauthor]{Corresponding author}
\ead{gopar@unizar.es}

\begin{abstract}

Random fluctuations of the shot-noise power in disordered graphene nanoribbons are studied. In particular, we calculate the distribution of 
the shot noise of nanoribbons with zigzag and armchair edge terminations. We show that the shot noise statistics is different for each type of 
these two graphene structures, which is 
 a consequence of presence of  different electron localizations: while  in zigzag nanoribbons electronic edge states are Anderson localized,   in 
armchair nanoribbons edge states are absent,  but electrons are anomalously  localized. 
Our analytical results are verified by tight binding numerical simulations  with random hopping elements, i.e., off diagonal 
disorder,  which preserves the symmetry of the graphene sublattices.
\end{abstract}

\begin{keyword}
Shot noise power,  graphene disordered nanoribbons, electron localization, 
\end{keyword}

\end{frontmatter}


\section{Introduction}

It is widely recognized that time dependent current fluctuations due to the discreteness of 
the electrical charges--shot noise power--provide further physical information of an electronic system   than other transport quantities such as the 
conductance. For instance, the shot noise takes into account the Pauli principle and  it can reveal electron correlations.  
Markus B\"uttiker and collaborators  
recognized the importance of the shot noise for the understanding of the problem of quantum electron transport and  
made major contributions to this topic. 

The shot-noise power can be studied by using a scattering approach to quantum  transport. Within 
this framework,  B\"uttiker found that 
the shot-noise power spectrum $P$, in the zero frequency limit and zero temperature, can be written as \cite{lesovik, buttiker}
\begin{equation}
 P =2eV G_0 \sum_{n=1}^N T_n(1-T_n) ,
 \label{P}
\end{equation}
where $G_0(=2e^2/h)$ is the conductance quantum, $V$ the applied voltage, while the $T_n$'s are the transmission eigenvalues of the 
Hermitean matrix $tt^\dagger$, $t$ being the $N\times N$ transmission matrix. If there were no correlations among electrons, the 
shot noise is given by the Poisson value $P_P=2eVG_0\sum_n T_n$. 
The shot noise power  has been extensively studied both experimentally and theoretically in small electronic devices such as quantum 
wires and quantum dots, in which  quantum coherence is preserved. The literature on this topic is very extensive, we thus refer the 
reader to  the review articles Refs. \cite{Jong, Blanter}. 

In general, electron correlations reduce the shot noise respect to the case of fully uncorrelated electrons. 
The Fano factor $F$ measures that suppression of the shot noise and it is  defined by the ratio  
$F=\langle P \rangle/\langle P_P \rangle $, where the brackets indicate energy or ensemble average. Using random matrix
theory to quantum transport, it has been predicted that the Fano factor takes the value 1/3 
for disordered quantum wires in the diffusive regime limit \cite{Beenakker_Buttiker, Nagaev}, while for ballistic chaotic quantum dots  
$F=1/4$,  in the limit of large number of channels supported by the  leads  attached to the dots \cite{Rodolfo,finitechannels}. Thus, 
universal values of the shot noise suppression have been predicted 
for both transport regimes.

With respect to graphene, several electronic properties have been intensively studied since its discovery in 2004  and 
the shot  noise is not an exception (for a review, see \cite{Castro} and \cite{ Das}).  For pristine graphene structures whose lengths are  
shorter than their widths, it  has been predicted a 1/3 shot-noise 
suppression, at the Dirac point \cite{Titov}. This suppression value  coincides  with the Fano factor value for disordered 
normal-metal wires in the diffusive  regime. 

Although pristine graphene structures show many interesting electronic properties, in real graphene-based devices those properties may be 
affected by the presence of disorder \cite{Das, mucciolo,Areshkin,Bardarson,Xiong,Louis,Evaldsson, Rotter, Wakabayashi,Ioannis_Elias}. 
In general, different  sources of disorder can be present in graphene such as ripples, vacancies,  adatoms, or 
distortions of  the lattice  produced by interactions with the substrate \cite{gallagher,melinda}, however, even suspended graphene 
structures are not free of defects \cite{meyer,jarillo}. 

In particular, effects of the presence of disorder on the shot noise in 
graphene sheets have been experimentally and theoretically investigated \cite{Das, Danneau, Dicarlo, Pablo, Caio, Ralitsa}. It has 
been found that  the Fano factor 
is affected by the strength of the disorder, as well as  the length-to-width ratio of the graphene sample. Most of those studies, however, have 
been concentrated on wide geometries and models of disorder that break the chiral symmetry of the graphene sublattices.

\begin{figure}
\begin{center}
\includegraphics[width=0.8\columnwidth]{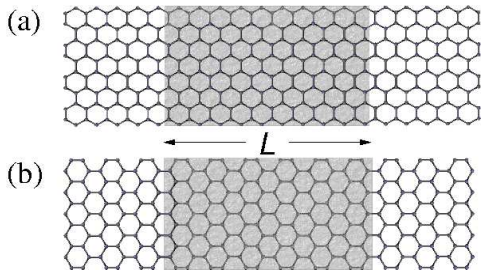}
  \caption{Disordered graphene nanoribbons (shaded areas) of length $L$  with zigzag (a) and armchair (b) edges with 
  perfect graphene leads (non-shaded areas) attached. The width $W$ of the nanoribbons has been fixed: $W=11 a/\sqrt{3}$ and 5$a$ for 
  zigzag and armchair nanoribbons  ($a$ being the lattice constant 
$a \simeq 2.46 \textup{\AA}$), respectively.  With those widths, the  attached perfect leads have gapless metallic band structures.}
  \label{figure_1}
  \end{center}
  \end{figure}

Here, we are interested in the properties  of the shot noise at the Dirac point in finite disordered graphene samples, i.e., 
disordered nanoribbons,  in which  the edge terminations as well as the symmetry of the graphene sublattices 
play
a crucial  role in the electronic properties. We thus investigate the 
random fluctuations of the shot noise power for  the  two different edge terminations: zigzag and armchair, Fig. \ref{figure_1}. For both  terminations, we 
consider 
the so-called  off-diagonal disorder (random hopping connecting the two graphene sublattices) in order to preserve the chiral symmetry 
of the graphene sublattices. This kind of short-range disorder might model  distortions like ripples  in the graphene lattice. 
Actually, experimental evidence of  short range disorder in graphene has been observed \cite{Tan,Jang,Zeuner}. Therefore,  
here we are interested in studying the effects of the nanoribbon edge terminations and the presence of disorder 
on the statistical properties of the shot noise power.
  
The most interesting properties of graphene are found at low energies, i.e., near the Dirac point  where  a linear dispersion relation 
holds and an analogy with 
relativistic massless particles has attracted much attention. Also, the  lattice symmetry of graphene or chiral symmetry, resulting
in a symmetric energy spectrum around the Fermi energy plays an important role in the description of the electronic properties; hence, we are 
interested in calculating 
statistical properties of the shot noise near the Dirac point. We recall that in clean graphene nanoribbons the band 
structure is determined by the edge termination and the width of the nanoribbons.  Here we consider disordered nanoribbons with 
attached perfect leads whose band structure is metallic. In the next section we introduce a 
statistical model to describe  the random fluctuations of the shot noise power, Eq. (\ref{P}).

  \section{Statistical Model}

The statistical properties of the shot noise power in disordered  graphene nanoribbons  can be analyzed through  the statistical 
properties of the transmission eigenvalues $T_n$, as we can see from Eq. (\ref{P}).  A well established theoretical 
framework to study the  transmission statistics of  disordered systems is the one-parameter scaling approach to localization \cite{Abrahams} and 
random matrix theory \cite{mello_book}.  Our statistical analysis of the shot noise is based on that 
scaling approach to  Anderson localization  and a 
recent extension to  the case of anomalous localization \cite{fernando_gopar, elias}. In general, the manner in which electrons are localized determines 
the statistical properties of quantum transport. Therefore, the statistics of the transmission, or conductance,  for standard (Anderson) and 
anomalous localizations is different. For instance, the average of the conductance of disordered systems in the presence of Anderson localization 
decays exponentially with the system length, whereas  the  conductance average has a power-law 
decay with the length for disorder systems with anomalous localization. Both kind of localizations have been investigated in disordered zigzag and 
armchair nanoribbons \cite{ioannis}.

On the one  hand,  near the Dirac point, only a single  transmission eigenvalue contributes to the electronic  transport. Therefore, 
since we are interested in the shot noise near the Dirac point, we only need to consider one transmission eigenvalue and  the shot-noise 
power in Eq. (\ref{P}) is reduced to $P=2eV G_0  T(1-T)$. For convenience, we introduced the dimensionless shot-noise power $S$ defined as:
\begin{equation}
 S=\frac{1}{2eV G_0}P=T(1-T) .
 \label{S}
\end{equation}
Thus, along this work we will be interested in describing the statistical properties  of the shot noise power $S$, as given by 
Eq. (\ref{S}). 

On the other hand, within the scaling approach to localization  and 
random matrix theory, the distribution of the transmission $P(T)$ is given by the solution of the so-called 
Mel'nikov's equation, which is an evolution equation of $P(T)$ with the length  of the disordered system \cite{Dorokhov,Melnikov, Abrikosov}. The 
exact solution of the Mel'nikov's equation is 
known in terms of quadratures. To simplify the calculations and provide  analytical expressions for shot noise distribution, we 
find convenient to consider an approximation to the exact  solution \cite{gopar-molina}:
 \begin{equation}
\label{pofT}
 P_s(T)=C \sqrt{ \frac{\mathrm{acosh} (1/\sqrt{T})}{T^3\sqrt{1-T}}}e^{-s^{-1}\mathrm{acosh}^2(1/\sqrt{T})} ,
\end{equation}
where $C$ is a normalization constant and $s$  is the ratio $L/l$, $L$ being the length of the system and $l$, the mean free path. The above 
 distribution, Eq. (\ref{pofT}), has been  verified in a number of numerical simulations for any practical value of the disorder strength  
 (measured by the value of $s$)   as well as in  microwave experiments \cite{Pena}. Additionally,  the value of the parameter $s$ can be obtained from
 the numerical or experimental data through the linear dependence of the average $\langle - \ln T \rangle$ with $L$: $\langle - \ln T \rangle =L/l (=s)$.
 
 We point out that the distribution given in Eq. (\ref{pofT}) is appropriate for  systems in which  the presence of   
 disorder leads to an  exponential localization of electron wavefunctions (Anderson localization) with the distance 
 $r$:  $| \psi|  \sim e^{-\gamma r}$, $\gamma$ being a constant. This exponential decay  has 
been experimentally and theoretically studied in several different  disordered systems \cite{phystoday}. 

The presence of disorder, however,
can lead to a different electron localization, or anomalous localization, in relation to the above-mentioned standard exponential decay. For 
example, in one-dimensional disordered systems, at the center of the band, the 
wavefunction decays as $| \psi|  \sim e^{-\gamma r^\alpha}$ with $\alpha=1/2$  \cite{soukoulis,inui}. For    
 disordered armchair nanoribbons, it has been also shown that electrons are anomalously localized ($\alpha=0.69$) \cite{ioannis}. 

Recently, it has been proposed a generalization of the single scaling approach  to describe the transmission through disordered systems 
with anomalous localization  \cite{elias}. In this case, the distribution of the transmission is given in terms of long-tailed 
probability density 
functions or L\'evy-type distributions $q_\alpha(x)$, where $\alpha$ is the power-tail exponent of the density function, i.e.,  for large $x$,
$q_\alpha (x) \sim 1/x^{1+\alpha}$.  The model is summarized by the following equation which gives the distribution of the transmission in 
terms of two quantities: the average $\langle \ln T \rangle $ and $\alpha$: 
 \begin{eqnarray}
\label{pofT_xi}
P_{\xi,\alpha}(T)=\int_0^\infty P_{\tilde{s}}(T) q_{\alpha}(z){\rm d}z  ,
\end{eqnarray} 
where we have defined $\xi =\langle - \ln T \rangle$. $P_{\tilde{s}(\alpha,\xi,z)}(T)$ is given by  Eq. (\ref{pofT}) with $s$ replaced by the 
function $\tilde{s}(\alpha,\xi,z)={\xi}/(2{z^\alpha I_\alpha)}$, where $I_\alpha =1/2 \int_{0}^{\infty} z^{-\alpha} q_{\alpha}dz $. The density function  
$q_{\alpha}(z)$ is part of a family of probability densities commonly known as  $\alpha$-stable distributions. 
We remark that only two quantities ($\xi$ and $\alpha$) determine the distribution of the transmission. Also, we point out that 
Eq. (\ref{pofT_xi}) predicts a nonlinear 
behavior of the logarithm of the transmission with the length $L$: $\langle \ln T \rangle \propto L^\alpha$, which is in contrast to the linear behavior 
expected 
for the standard Anderson localization. 
The   model summarized in Eq. (\ref{pofT_xi}) has been applied recently to describe the transmission of microwaves in disordered  waveguides \cite{prl_antonio}
 
 In the following  two sections, we will apply  Eqs. (\ref{pofT}) and (\ref{pofT_xi}) to calculate the distribution of the shot noise 
 power $S$,  Eq. (\ref{S}), in zigzag and armchair nanoribbons, respectively. The theoretical predictions will be compared with numerical 
 simulations using  a standard tight-binding Hamiltonian model: 
\begin{equation}
\label{tight}
  H= \sum_{<i,j>}t_{i,j} (c_{i}^\dagger c_{j} +c_{j}^\dagger c_{i} ) , 
\end{equation}
where $i$ and $j$ are nearest neighbors and $c_i^\dagger$ ($c_i$) is the creation (annihilation) operator for spinless fermions, while the 
hopping elements $t_{i,j}$  between the two graphene sublattices are randomly obtained  from the distribution 
$p(t)=1/wt$ with $\exp(-w/2) \leq t \leq \exp (w/2) $, $w$ being the strength of the disorder which is fixed to 1. This  short-range 
disorder models random distortions in graphene sheets without breaking the symmetry of the graphene lattice. The  
numerical simulations were performed  near the Dirac point: $E=10^{-6}$ (in units of the hopping energy of the perfect leads) for  zigzag and 
armchair  nanoribbons of  widths $W=11a/ \sqrt{3}$ and  $W=5 a$ ($a$ being the lattice constant 
$a \simeq 2.46 \textup{\AA}$), respectively. The transmission eigenvalues in Eq. (\ref{S}) were calculated by 
using a recursive Green's function and, for all cases shown in this work,  the shot 
noise statistics were collected  from an ensemble of $2 \times 10^4$  disorder realizations.
 
 \section{zigzag nanoribbons: shot noise via edge states} 

  Firstly, we  consider  the case of disordered zigzag nanoribbons, Fig. \ref{figure_1}(a). 
 In pristine zigzag nanoribbons, it is well known the existence  of edge states, which are perfectly transmitted.   In the presence of 
 disorder, those states remain at the border of the nanoribbons, but they are (Anderson) localized \cite{ioannis}. On the other hand, since near 
 the Fermi energy  only a  single channel contributes to the transmission, we can use the distribution function  in Eq. (\ref{pofT}) to calculate 
 the distribution of the 
 shot-noise power given by $P(S)=\langle \delta(S-T(1-T))\rangle$.  Thus, using Eqs. (\ref{S}) and (\ref{pofT}), it is  straightforward but 
 lengthy to perform the average indicated with brackets. The final expression for the distribution of the shot noise power is  
  \begin{eqnarray}
\label{pofS}
 & &P_s(S)= \frac{C}{\sqrt{1+4S}}\nonumber \\
 &&\times\left\{
  \sqrt{ \frac{\mathrm{acosh} \left(1/\sqrt{S_+}\right)}{{S^3_+}\sqrt{1-S_+}}}e^{-s^{-1}\mathrm{acosh}^2\left(1/\sqrt{S_+}\right)}\right. 
  \nonumber \\
 &&\left. +\sqrt{ \frac{\mathrm{acosh}\left (1/\sqrt{S_-}\right)}{{S^3_-}\sqrt{1-S_-}}}e^{-s^{-1}\mathrm{acosh}^2\left(1/\sqrt{S_-}\right)}
 \right\},\nonumber\\
\end{eqnarray}
where $C$ is a normalization constant and   $S_\pm=(1 \pm \sqrt{1+4S})/2$. We notice that the 
statistical properties of the shot noise are determined by the single parameter $s$, which can be extracted from experimental or numerical data, as we 
have pointed out. 

\begin{figure}
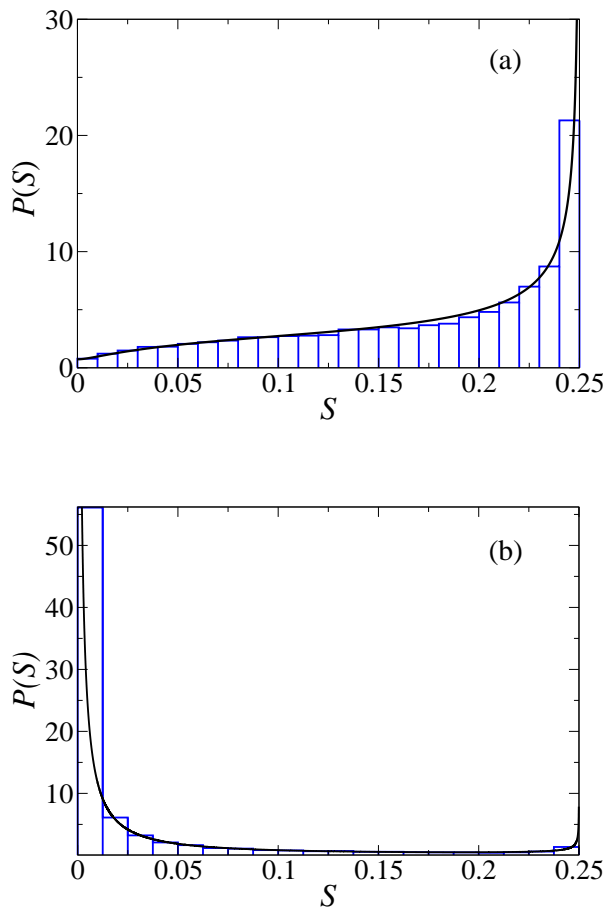

\begin{center}
\includegraphics[width=\columnwidth,angle=0]{fig_2a_physica.eps}
\includegraphics[width=\columnwidth]{fig_2b_physica.eps}\\
  \caption{Distribution of the shot noise power for disordered zigzag nanoribbons. Panels (a) and (b) show 
  the numerical (histograms) and theoretical distribution (solid line) for disordered nanoribbons with dimensionless conductance  average 
  $\langle G \rangle =0.46$ ($s=1.03$) and $\langle G \rangle =0.043$ ($s=6.21$), respectively.  The value of the Fano factor is 0.37 for panel (a), while 
  $F=0.6$ for panel (b). A good agreement is seen between theory and numerical simulations.}
  \label{figure_2}
  \end{center}
  \end{figure}

In order to verify the above result,  we compare the theoretical distribution, Eq. (\ref{pofS}),  with the tight binding numerical simulations,  described 
previously. In  Fig. \ref{figure_2}, the 
numerical distributions (histograms) and $P_s (S)$ (solid line) are 
compared for two different strength of disorder, measured by the parameter $s$. The  
value of $s$ is obtained from the numerical data and  it is plugged into Eq. (\ref{pofS}). Figs. \ref{figure_2}(a) and \ref{figure_2}(b) correspond to 
zigzag nanoribbons 
which are characterized by the average dimensionless conductance $\langle  G \rangle  =0.46 $ and $\langle G \rangle =0.04$, respectively. The Fano 
factor values for 
those cases are $F=0.37$ for Fig. \ref{figure_2}(a), while  $F=0.6$ for the case shown in Fig. \ref{figure_2}(b). 
We can observe a good agreement between the model [Eq. (\ref{pofS})] 
and the numerical simulations. 
  
 \section{Armchair nanoribbons: shot noise via anomalously localized states}
 
 We now consider the case of nanoribbons with armchair  terminations, Fig. \ref{figure_1}(b). As we have mentioned, evidence of the presence of 
 anomalous localization has been 
 found  when the hopping elements of the tight binding Hamiltonian [Eq. (\ref{tight})]  have random 
 fluctuations, which  may model random 
 distortions in the graphene lattice.  As in the case of zigzag nanoribbons, near the Fermi energy, only a single channel contributes to the electronic transport
 and the shot noise distribution can be obtained by performing  the average  $\langle \delta(S-T(1-T))\rangle$. In  
 the present case, however, the average indicated with brackets is performed with the transmission distribution given by Eq. (\ref{pofT_xi}).  
  The $\alpha$ parameter of $q_\alpha(z)$ in Eq.  (\ref{pofT_xi}) that characterize 
 the anomalous localization of our disordered armchair nanoribbons has been numerically obtained ($\alpha= 0.69$) in \cite{ioannis}. This value 
 of $\alpha$ does not depend on the length, width, and strength of the disorder, according to the numerical simulations. On the other hand,  
 unfortunately, there is no a close analytical 
 expression for the L\'evy-type distribution $q_\alpha(z)$  with $\alpha=0.69$; thus,  we express the shot noise distribution for armchair nanoribbons 
 in terms of quadratures:
 \begin{equation}
 \label{pofSstilde}
  P_{\xi, \alpha} (S)=\int \left[ P_{\tilde{s}}(S_+) + P_{\tilde{s}}(S_-)\right] q_\alpha(z)dz ,
 \end{equation}
 where $ P_{\tilde{s}} (S)$ is given by Eq. (\ref{pofS}) with $s$ replaced by $\tilde{s}={\xi}/(2{z^\alpha I_\alpha)}$. As we have defined previously:  
 $\xi =\langle - \ln T \rangle$, $I_\alpha =1/2 \int_{0}^{\infty} z^{-\alpha} q_{\alpha}dz $, and $S_\pm=(1 \pm \sqrt{1+4S})/2$. 
 
\begin{figure}
\begin{center}
\includegraphics[width=\columnwidth]{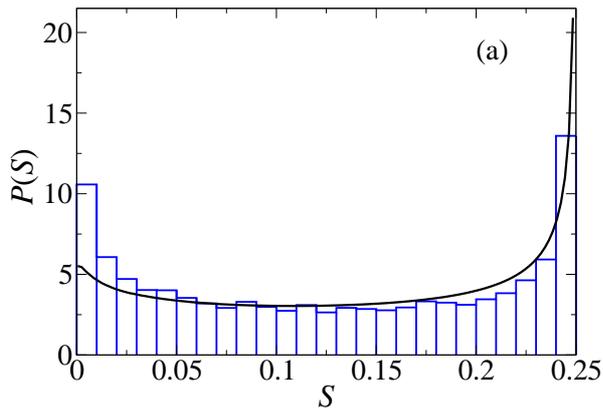}
\includegraphics[width=\columnwidth]{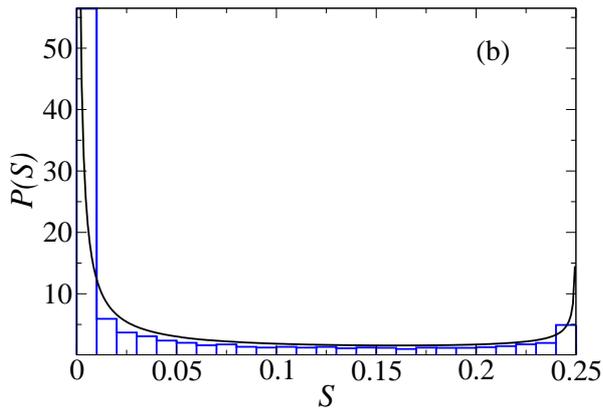}
  \caption{Shot noise distribution for disordered armchair nanoribbons. Panels (a) and (b) show 
  the numerical (histograms) and theoretical distribution (solid line) for disordered nanoribbons with transmission average 
  $\langle G \rangle =0.46$ ($\langle - \ln G \rangle=1.37$)  and $\langle G \rangle =0.15$ ($\langle - \ln G \rangle=6.37$) for panels (a) and (b), respectively.  The value of the Fano factor is 0.27 for panel (a), while 
  $F=0.35$ for panel (b). We can observe that the theoretical predictions reproduce the trend of the numerical distributions.}
  \label{figure_3}
  \end{center}
  \end{figure}

 We now compare our 
 expression in Eq. (\ref{pofSstilde}) with numerical simulations of disordered armchair nanoribbons.  In Fig. \ref{figure_3} we show the 
 numerical (histogram) and theoretical (solid line) shot  noise distributions for two different 
 values of $\xi$. The values of $\xi$ are extracted from the numerical data and they are plugged into Eq. (\ref{pofSstilde}).  We may also 
 characterized the nanoribbons by the average dimensionless conductance: in Fig. \ref{figure_3}(a),  $\langle G \rangle =0.46$, while in 
 Fig. \ref{figure_3}(b)  $\langle G \rangle =0.15$.  The values of the Fano factor of these two cases are $ F=0.27$  for panel (a), while  
 $F=0.35$ for panel (b).  We can observe 
 that our model describe correctly the trend of the shot noise distributions. As in the previous case of disordered zigzag nanoribbons, we point out 
 that  no parameters have been adjusted in our theoretical results. 
 
\begin{figure}
 \begin{center}
\includegraphics[width=\columnwidth]{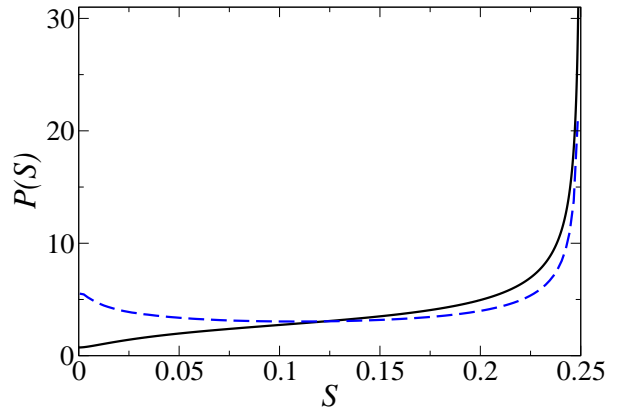}
  \caption{Shot noise distributions for zigzag (solid line) and armchair (dashed line) nanoribbons with the same average dimensionless conductance  
  ($\langle G \rangle=0.46$). Stronger transmission fluctuations due to the presence of anomalous localization in armchair nanoribbons 
  increase the probability of smaller values of the shot noise power, in relation to the zigzag nanoribbons.}
  \label{figure_4}
  \end{center}
  \end{figure}

 Finally, we would like  to contrast  the statistics of the shot noise power for both type nanoribbon terminations; hence, in Fig. \ref{figure_4} we show the 
 shot noise distributions for both armchair (dashed line) and zigzag  (solid line) nanoribbons. For both cases, we     
  have chosen  nanoribbons with the same  average of the dimensionless conductance ($\langle G \rangle =0.46$), shown in Figs. \ref{figure_2}(a) and 
  \ref{figure_3}(a). For  this value of $\langle G \rangle$, 
  the Fano factor values are: $F=0.37$ for zigzag nanoribbons, while $F=0.27$ for armchair nanoribbons.  From Fig.  \ref{figure_4}, 
  the   main differences in the distributions are seen at small values of the shot noise. We can also observe that the probability  of having small 
  values of the shot noise is larger   in armchair 
 nanoribbons than in zigzag nanoribbons. 
 This can be explained  by  the larger fluctuations in the transmission eigenvalues in armchair nanoribbons, as a consequence of the 
 presence of anomalous localization, which lead to large fluctuations of the shot noise power.  Thus, Fig. \ref{figure_4} shows an example of the effects 
 of the edge termination on the statistics of the  shot noise power.

\section{Summary and Conclusions}
We have investigated
the statistics of the random fluctuations of the shot noise power in disordered graphene nanoribbons with zigzag and armchair 
edge terminations, near the Dirac point.  Within a scattering approach developed  by B\"uttiker and collaborators,  the shot noise can be 
written in terms of the transmission eigenvalues. We thus apply a random matrix theory of quantum transport  to calculated the complete distribution of
the shot noise power for both type of nanoribbon edge terminations. 

Our theoretical model predicts different statistical properties of the shot noise
for zigzag and armchair nanoribbons. Those differences come from the fact that electronic edge states  in zigzag nanoribbons are exponentially 
localized in space, i.e., electrons are (Anderson) localized, while in  armchair nanoribbons, electrons are anomalously localized  or delocalized, in 
relation to the case of  Anderson localization. Anomalous localization produces stronger random  fluctuations of the transmission which lead to a 
different statistics of the shot noise. We point out that while for zigzag nanoribbons the shot noise distribution 
depends on a single parameter, for armchair nanoribbons the  shot noise distribution is determined by two parameters. Those parameters are not free 
in our model in the sense that their values are  obtained from the  numerical simulations, or experimental data, and they are used as an input in the analytical expressions.  
Our theoretical predictions have been verified by numerical simulations of graphene nanoribbons  with off-diagonal 
disorder, which preserve the chiral symmetry of the graphene sublattices.  We have found that the value of the Fano factor  is not 
universal but it depends on the strength of the disorder, which is in contrast to the known universal value 1/3  for short and wide clean 
graphene structures.

Therefore, to conclude, we have shown  that nanoribbon edge  terminations as well as  the presence of disorder  play are relevant role in the 
properties of the shot noise power in disordered graphene nanoribbons.

\section{Acknowledgements}
I thank Ilias  Amanatidis and Ioannis Kleftogiannis for discussions and numerical support,  as well as the hospitality of the 
National Center for Theoretical Science, Taiwan, where those fruitful  discussions  took place. This work was partially supported 
by  MINECO (Spain) under Project  No. FIS2012-35719-C02-02.

\end{document}